\documentstyle[12pt,lnfprep,epsfig,amsmath]{article}
\newcommand{\be}{\begin{equation}}
\newcommand{\ee}{\end{equation}}
\newcommand{\bea}{\begin{equation} \begin{array}{c}}
\newcommand{\eea}{ \end{array} \end{equation}}
\newcommand{\bad}{\begin{array}{ccc}}
\newcommand{\ea}{\end{array}}
\def\simge{\mathrel{%
      \rlap{\raise 0.511ex \hbox{$>$}}{\lower 0.511ex \hbox{$\sim$}}}}
\def\simle{\mathrel{
      \rlap{\raise 0.511ex \hbox{$<$}}{\lower 0.511ex \hbox{$\sim$}}}}
\newcommand{\Header}{
     \begin{tabular}{rl}
     \hspace{-.4cm}\includegraphics{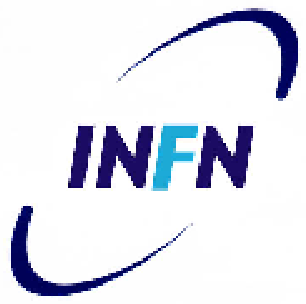} &
       \renewcommand{\arraystretch}{0.5}
       \begin{tabular}{r}
         {\hspace{1cm}~\LARGE\sffamily LABORATORI~ NAZIONALI~ DI~ FRASCATI}\\
         \\
         {\Large\sffamily SIS-Pubblicazioni}\\
       \end{tabular}
       \renewcommand{\arraystretch}{1}
     \end{tabular}
     \vskip 1cm
     \begin{flushright}
     \renewcommand{\arraystretch}{0.5}
       \begin{tabular}{r}
         {LNF-04/07}\\    
       \end{tabular}
     \end{flushright}
     \renewcommand{\arraystretch}{1}
     \vskip 1 cm
     }

\begin{document}
\begin{titlepage}
\title{
     \Header
     {\large \bf NEUTRINO OSCILLATION STUDIES WITH LASER-DRIVEN
BEAM DUMP FACILITIES}
}
\author{S.V.~Bulanov$^{1,5}$, T.~Esirkepov$^{1,6}$, P.~Migliozzi$^{2}$,
F.~Pegoraro$^{3}$, T.~Tajima$^{1}$, F.~Terranova$^{4}$ \\
{\it ${}^{1)}$ Kansai Research Establishment,
JAERI, Japan.}
     \\
{\it ${}^{2)}$INFN, Sez. di Napoli, Napoli, Italy}
     \\
{\it ${}^{3)}$Dip. di Fisica, Univ. di Pisa and INFM, Pisa, Italy}
     \\
{\it ${}^{4)}$INFN, Laboratori Nazionali di Frascati, Frascati, Italy}
     \\
{\it ${}^{5)}$ A.~M.~Prokhorov Institute of RAS, Moscow, Russia}
     \\
{\it ${}^{6)}$ Moscow Institute of Physics and Technology, Dolgoprudny,
Russia. }
}
\maketitle
\baselineskip=14pt

\begin{abstract}
A new mechanism is suggested for efficient proton acceleration in the
GeV energy range; applications to non-conventional high intensity
proton drivers and, hence, to low-energy (10-200~MeV) neutrino sources
are discussed. In particular we investigate possible uses to explore
subdominant $\bar{\nu}_\mu \rightarrow \bar{\nu}_e$ oscillations at
the atmospheric scale and their CP conjugate. We emphasize the
opportunity to develop these facilities in conjunction with projects
for inertial confined nuclear fusion and neutron spallation sources.
\end{abstract}

\vspace*{\stretch{2}}
\begin{flushleft}
\end{flushleft}
\begin{center}

\end{center}
\end{titlepage}
\pagestyle{plain}
\setcounter{page}2
\baselineskip=17pt

\section{Introduction}
\label{introduction}

In the light of the strong evidence~\cite{evidence_osc} for neutrino
oscillations coming from atmospheric and solar
neutrino experiments, recently corroborated by
reactor and accelerator results, a very peculiar texture of the
leptonic mixing matrix is emerging.  Current results point towards two
hierarchical mass scale differences\footnote{We assume here only three
active neutrinos with masses $m_1$, $m_2$ and $m_3$; hence, only two
independent mass scale differences exist ($\Delta m^2_{12}\equiv m_2^2
- m_1^2$ and $\Delta m^2_{23}\equiv m_3^2 - m_2^2$) since $\Delta
m^2_{13}=\Delta m^2_{23}+\Delta m^2_{12}$.}  
($\Delta m^2_{12} \ll
|\Delta m^2_{13}| \simeq |\Delta m^2_{23}|$) driving, respectively,
the oscillations at the ``solar'' and ``atmospheric'' scales. The
corresponding mixing angles are large but the interference between the
two scales (sub-dominant $\nu_\mu \rightarrow \nu_e$ oscillations at
$E/L \sim |\Delta m^2_{23}|$, $L$ and $E$ being the neutrino
path-length and energy) has never been
observed~\cite{CHOOZ,paloverde}. The leptonic mixing matrix
(Pontecorvo-Maki-Nakagawa-Sakata, PMNS~\cite{PMNS}) is usually
parametrized~\cite{Hagiwara:fs} as:

\bea 
U(\theta_{12},\theta_{13},\theta_{23},\delta)=   \\ 
\left( \bad 
c_{12} c_{13} & s_{12} c_{13} & s_{13}  \\[0.2cm] 
-s_{12} c_{23} - c_{12} s_{23} s_{13} e^{i \delta} 
& c_{12} c_{23} - s_{12} s_{23} s_{13} e^{i \delta} 
& s_{23} c_{13} e^{i \delta}\\[0.2cm] 
s_{12} s_{23} - c_{12} c_{23} s_{13} e^{i \delta} & 
- c_{12} s_{23} - s_{12} c_{23} s_{13} e^{i \delta} 
& c_{23} c_{13}e^{i \delta}\\ 
               \ea   \right) 
\eea

\noindent with $s_{ij}=\sin\theta_{ij}$ and $c_{ij}=\cos\theta_{ij}$;
current data suggest~\cite{Maltoni:2003da} at 90\% confidence level
$35^\circ<\theta_{23}<55^\circ$ , $\theta_{12}=32.5^\circ
\pm2.4^\circ$ and small values for $\theta_{13}$ ($\simle 10^\circ$)
i.e. support a ``bi-large'' PMNS. In fact, in the limit $\theta_{13}
\rightarrow 0$ the matrix becomes real-valued and the complex CP
violating phase turns out to be unobservable. It follows that the
possibility to determine experimentally the $(1,3)$ sector of PMNS -
i.e. the off-diagonal factor $U_{e3} \equiv s_{13}e^{-i\delta}$ - and,
in particular, the Dirac phase $\delta$ critically depends on the size
of $\theta_{13}$.  No theoretical inputs are available to constrain
the size of $\theta_{13}$ and $\delta$ in a convincing manner, so that
its experimental determination is mandatory.  Such determination can
be carried out at accelerators either measuring the size of the
subdominant $\nu_\mu \rightarrow \nu_e$ ($\bar{\nu}_\mu \rightarrow
\bar{\nu}_e$) oscillation probability at the atmospheric scale or its
T-conjugate $\nu_e \rightarrow \nu_\mu$ ($\bar{\nu}_e \rightarrow
\bar{\nu}_\mu$). This measure will likely be the most challenging task
of future long-baseline neutrino oscillation experiments and, for
$\theta_{13}$ values significantly smaller than current limits,
traditional accelerating techniques will be unable to provide the
requested intensity and purity. Therefore, interest for unconventional
neutrino sources has steadily grown in recent years, bringing e.g.  to
the proposals of the Neutrino Factories~\cite{Geer:1997iz} or the
Beta-beams~\cite{Zucchelli:sa}.

Along this line, in this paper we consider a wide synergic scenario
between nuclear and neutrino physics programs connecting the long term
development of facilities for laser-driven inertial confinement fusion
(ICF) and the possibility to obtain an ultra-intense low-energy
(1-2~GeV) proton driver for neutrino studies and spallation neutron
sources (see Appendix). Such a link is made possible if an efficient
laser-based acceleration mechanism is available in the near
future. This mechanism is described in details in
Sec.~\ref{sec:laser}. The connection with neutrino physics results
from the following consideration.  Sources of neutrinos with energy
beyond the $\mu$ production threshold in $\nu_\mu$ charged-currents
(CC) interactions are not strictly necessary to explore the magnitude
and phase of $U_{e3}$. In principle, a high intensity source of
$\nu_\mu$ and $\bar{\nu}_\mu$ with energy of the order of a few tens
of MeV would suffice to search for $\nu_\mu \rightarrow \nu_e$
appearance and its CP-conjugate at baselines of $L\sim 10$~km.  The
neutrinos come from pion and muon decays at rest (DAR) and the pions
can be produced by a high current proton beam dump facility. These
beams, if available with proper intensity would provide {\it
simultaneously} a source for $\nu_\mu \rightarrow \nu_e$ oscillations
through the $\pi^+ \rightarrow \mu^+ \nu_\mu$ DAR chain and a source
for $\bar{\nu}_\mu \rightarrow \bar{\nu}_e$ through the subsequent
$\mu^+ \rightarrow e^+ \bar{\nu}_\mu \nu_e $ decay. Moreover,
differently from Superbeams~\cite{JHF,BNL}, the intrinsic $\nu_e$
($\bar{\nu}_e$) beam contamination can be kept easily below 0.1\% (see
Sec.~\ref{sec:nubeams}), as the facility can be operated below the $K$
production threshold.

The acceleration of protons up to several tens of MeV by the
interaction of ultra-intense laser beams with solid targets has been
recently reported by several experiments \cite{ref:laserexp}. This
process will probably open up a wealth of applications: radioisotope
production~\cite{ref:spencer}, proton probing~\cite{ref:borghesi} and
oncological hadron-therapy~\cite{ref:bulanov_plb} have been discussed.
Moreover, the laser-driven acceleration mechanism is a natural
candidate for fast beam injection into conventional
accelerators~\cite{ref:krushelnik}.  Part of these research programs
might be carried out already with present laser technologies, provided
that a suitable repetition rate, reproducibility and an improved beam
quality become available. 
On the other hand, future
developments towards higher laser intensities~\cite{tajima,Es} would
allow particle acceleration for High Energy Physics applications.

In fact, laser-driven acceleration mechanisms mainly favor
applications that profit of the significant beam intensity without
imposing strong constraints on the beam quality (energy spread and
emittance) and particle energy. 
The possibility of using the well-controlled time structure of
laser-induced $\nu$ beams to test the KARMEN time anomaly or
increasing the laser intensity to overcome the kinematical threshold
for muon production ($\nu_\mu$ disappearance tests) has already been
considered in~\cite{pakhomov}. The pion generation by the laser
accelerated ions has also been discussed in~\cite{bychenkov}.
Similarly, in this paper we suggest that the above-mentioned technique
could be implemented to overcome the present limitations of proton
dump facilities (Sec.~\ref{sec:laser}) and, particularly, high
intensity DAR neutrino beams (Sec.~\ref{sec:nubeams}). We investigate
its capability to clarify the (1,3) sector of the PMNS matrix, with
emphasis on the size of $\theta_{13}$
(Sec.~\ref{sec:channel_detector}-\ref{sec:theta_13}), and the main
technological challenges for a new generation of beam dump facilities.

\section{Laser ion accelerators}
\label{sec:laser}

The classical mechanism of ion acceleration through the interaction of
a laser pulse with a plasma target is a direct consequence of the
electron acceleration. Due to the smaller electron mass, the energy of
the laser light is first transformed into electron kinetic energy. The
resulting displacement of the electrons and the modification of their
density lead to the formation of a region with a strong electric
charge separation. This causes an intense electric field which
eventually accelerates the ions. In the simplest case of a
one-dimensional geometry, when a transversally wide laser pulse
interacts with a thin foil, the ponderomotive pressure of the laser
pulse shifts the electrons with respect to ions, forming a strong
electric field layer between the two species. An electric field also
can be formed due to charge separation if the laser radiation
accelerates a relatively large portion of the electrons, expelling
them almost isotropically.  In this case the fast electrons leave the
targets and the heavier ions remain at rest forming extended regions
of positive electric charge.  Then the ions with non-compensated
electric charge expand and acquire a kinetic energy. This corresponds
to the so called ``Coulomb explosion'' regime. In configurations with
more than one dimensions, different effects come into play such as the
finite size of the waist of the laser beam, the transverse
filamentation instability of a wide electromagnetic packet in a
plasma, and the transverse modulations of the electron and ion
layers. These effects usually reduce the energy of the fast ions
and/or the efficiency of the energy transformation compared to the one
obtained within the framework of the one-dimensional approximation.

However, the process of ion acceleration exhibits new properties in
the regime where the radiation pressure of the electromagnetic wave
plays a dominant role in the laser-foil interaction, as demonstrated
in Refs. \cite{Es,BEKT}. In this regime electrons gain their energy
due to the radiation pressure in a way that, qualitatively, resembles
the ion acceleration mechanism proposed by Veksler~\cite{Veksler} in
1956.  Here, the ion component moves forward with almost the same
velocity as the average longitudinal velocity of the electron
component, hence with a kinetic energy well above that of the electron
component. The ion acceleration appears to be due to the radiation
pressure of the laser light on the electron component with the
momentum being transferred to the ions through the electric field
arising from charge separation.  This mechanism of ion acceleration
can be called the Radiation Pressure Dominant (RPD) mechanism. In
contrast to the schemes previously discussed in the literature (see
e.g. \cite{R1,R2,R3}) here the ion beam generation is highly
efficient, and, as we will see later on, the ion energy per nucleon is
proportional to the laser pulse energy.

\subsection{Radiation Pressure-dominated (RPD) regime: 1D analytical 
description}
\label{sec:1d}

The acceleration mechanism can be explained as follows. The
accelerated foil, which consists of the electron and proton layers,
can be regarded as a relativistic plasma mirror co-propagating with
the laser pulse. Assume that the laser pulse is perfectly reflected
from this mirror. As a result of the reflection at the co-propagating
relativistic mirror, the frequency of the electromagnetic wave
decreases by the factor of $\left( 1-v/c\right) /$ $\left(
1+v/c\right) \approx 1/4\gamma ^{2}$, where $v$ is the mirror velocity
and $\gamma =\left( 1-v^{2}/c^{2}\right) ^{-1/2}$ is the
Lorentz-factor of the plasma mirror\footnote{To ease comparison with
existing literature, we use here the Gaussian CGS system of unit
($c\ne 1$). We resume the usual $c=1$ system in the subsequent
sections.}.

\begin{figure}
\begin{center}
\resizebox{0.7\textwidth}{!}{
\includegraphics{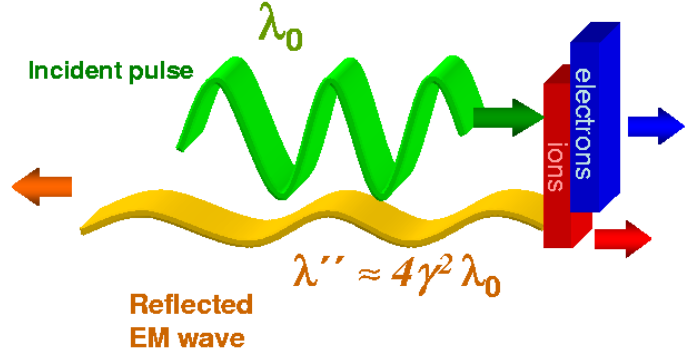} } 
\end{center}
\caption{Interaction of the e.m. wave  with the co-propagating
proton-electron mirror.}
\label{fig:mirror}
\end{figure}

Before the reflection, in the laboratory reference frame, the incident
laser pulse energy ${\cal E}_{las,in}$ is proportional to $
E_{0}^{2}L_{las,in}$, where $E_{0}$ is the electric field amplitude
and $L_{las,in}$ is the incident pulse length. After the reflection
the pulse energy becomes much lower: ${\cal E}_{las,ref}$ $\propto$
$E_{las,ref}^{2}$ $L_{las,ref}$ $\approx E_{0}^{2}L_{las,in}/4\gamma
^{2}$. The length of the reflected pulse is longer by a factor
$4\gamma ^{2}$, and the electric field is smaller by a factor
$4\gamma ^{2}$. Hence the plasma mirror acquires the energy
$(1-1/4\gamma ^{2}){\cal E}_{las,in}$ from the laser. In this stage
the radiation pressure of the light accelerates the plasma slab
(electrons and protons). As discussed above, the radiation momentum is
transferred to the protons through the charge separation field and the
kinetic energy of the protons is much greater than that of the
electrons. This scenario is illustrated in Fig.~\ref{fig:mirror}, where
a sketch of the e.m. wave interaction with the co-propagating
proton-electron slab is presented. 

It is possible to estimate the proton maximum energy and the
acceleration efficiency using the model of the flat foil driven by the
e.m. radiation pressure, as described above. The radiation pressure is
given by:
\begin{equation}  \label{P}
P=\frac{E_{0}^{\prime 2}}{2\pi }=\left( \frac{\omega ^{\prime }}
{\omega }\right) ^{2}\frac{E_{0}^{2}}{2\pi
}=\frac{E_{0}^{2}}{2\pi }\left( \frac{c-v}{c+v}\right) ,
\end{equation}
where $v=dx/dt$ is the foil instantaneous velocity, i.e.
\begin{equation}  \label{vel}
\frac{dx}{dt}=c\frac{p}{(m_{p}^{2}c^{2}+p^{2})^{1/2}}.
\end{equation}
Here primed (unprimed) quantities refer to the moving (laboratory) reference
frame respectively. In a quasi-one-dimensional geometry, the laser electric
field at the foil location $x(t)$ depends on time as $E'_{0}=E_{0}(t-x(t)/c)$.

Since the expression of the radiation pressure is the same in both
frames,  we can write the equation of motion of the foil as
\begin{equation}
\frac{dp}{dt}=\frac{E_{0}^{2}(t-x(t)/c)}{2\pi n_{0}l_{0}}
\frac{(m_{p}^{2}c^{2}+p^{2})^{1/2}-p}
{(m_{p}^{2}c^{2}+p^{2})^{1/2}+p}\,,
\label{eq-p}
\end{equation}
where $p$ is the momentum of the proton representing the foil, $l_0$
and $n_0$ are the thickness and initial proton density of the
foil. In this approximation we neglect the heating of the proton
fluid.

In the simplest case, when the laser pulse is assumed to be long
enough and with a homogeneous amplitude, we can consider the electric
field $E_0$ in Eq. (\ref{eq-p}) to be constant.  Its solution $p(t)$
is an algebraic function of time $t$. For the initial condition $p=0$
at $t=0$ it can be written in the implicit form
\begin{equation}
\frac{2p^{3}+2(m_{p}^{2}c^{2}+p^{2})^{3/2}}{3m_{p}^{2}c^{2}}+p-
\frac{2}{3} m_{p}c=\frac{E_{0}^{2}}{2\pi n_{0}l_{0}}t.  \label{rad-press}
\end{equation}
At the initial stage, for $t\ll 2\pi n_{0}l_{0}m_{p}c/E_{0}^{2}$, the proton
momentum is a linear function of time:
\begin{equation}
p\approx (E_{0}^{2}/2\pi n_{0}l_{0})t.  \label{rad-press1}
\end{equation}
As $t\rightarrow \infty $, the dependence of the accelerated proton momentum
on time changes asymptotically to
\begin{equation}
p\approx {\cal E}_{p\mathrm{kin}}/c\approx m_{p}c(3E_{0}^{2}t/8\pi
n_{0}l_{0}m_{p}c)^{1/3}.  \label{rad-press2}
\end{equation}
We notice here the obvious analogy between the proton motion regime
described by the expression (\ref{rad-press}) and the solution of the
problem of the acceleration of a charged particle under the radiation
pressure of the electromagnetic wave. This analogy can be clearly seen
by comparing Eq. (\ref{rad-press}) with the expression for the
velocity of an accelerated electron, which, in this limit, can be cast
in the form $\bar{W}\sigma_{T}~t/m_{e}c\approx
(2/3)(1-v^{2}/c^{2})^{3/2}$, with $c \bar{W}=cE_{0}^{2}/4\pi $ the
wave intensity (see Ref.~\cite{TeorPol}), $\sigma _{T}=8\pi
r_{e}^{2}/3$ the Thomson cross section and $r_{e}=e^{2}/m_{e}c^{2}$
the classical electron radius. In our case  $2/n_{0}l_{0}$ plays
the role of the effective cross section of the scattering of the
electromagnetic wave.  This analogy further underlines the similarity
between the RPD mechanism of ion acceleration discussed here and the
Veksler mechanism mentioned above.

To find an upper limit to the proton energy acquired during the
interaction with a laser pulse of finite duration, we must include the
dependence of the laser electromagnetic field on time $t$ and on the
coordinate $x$.  Because of the foil motion, the interaction time can
be much longer that the laser pulse duration $\tau _{las,in}$. We
introduce the phase of the wave
\begin{equation}
\psi =\omega _{0}(t-x(t)/c),  \label{psi}
\end{equation}
as a new variable, $\omega_0$ being the incoming laser
frequency. Using Eq.(\ref{vel}) together with Eq.(\ref{psi}),
we cast Eq.(\ref{eq-p}) for the particle momentum to the form
\begin{equation}
\frac{dp}{d\psi }=\frac{E_{0}^{2}(\psi )}{2\pi \omega _{0}n_{0}l_{0}}
\frac{(m_{p}^{2}c^{2}+p^{2})^{1/2}}{(m_{p}^{2}c^{2}+p^{2})^{1/2}+p}\,.
\end{equation}
Its solution reads
\begin{equation}
p=m_{p}c\frac{w(w+1)}{(w+1/2)}\,\,,  \label{eq-p-W}
\end{equation}
where $w$ is a function of $\psi $:
\begin{equation}
w(\psi )=\int_{-\infty }^{\psi }\frac{E_{0}^{2}(s)}{4\pi \omega
_{0}n_{0}l_{0}m_{p}c}ds\,.  \label{eq-w-psi}
\end{equation}
\noindent Using Eqs.(\ref{psi}) and (\ref{eq-p-W}) we find the 
dependence of
the foil coordinate on time and  write the  equations for $t$ and $x$ as
functions of the variable $\psi$ in the form 
\begin{eqnarray}\label{AA1}
\frac{dt}{d\psi } & = &
\frac{1}{\omega _{0}}\left( \frac{c}{c-v}\right) =\frac{1}
{\omega
_{0}}\frac{(m_{p}^{2}c^{2}+p^{2})^{1/2}}{(m_{p}^{2}c^{2}+p^{2})^{1/2}-p} 
\nonumber \\
& = & \frac{1}{\omega _{0}}\left[
1+{2w(w+1)}\right]
\end{eqnarray}
and
\begin{equation}\label{AA2}
\frac{dx}{d\psi }=\frac{dx}{dt}\frac{dt}{d\psi }=\frac{c}{\omega_{0}}2w(w+1).
\end{equation}
For a constant amplitude laser pulse with
$E_{0}^{2}(\psi)=E_{0}^{2}\theta (\psi )$, where $\theta (\psi )=0$ 
for $\psi <0$
and
$\theta (\psi )=1$ for $\psi >0$ is the unit step function,
we have $w(\psi )=w_{0}\theta (\psi )\psi $,
with $w_{0}=E_{0}^{2}/(4\pi \omega _{0}n_{0}l_{0}m_{p}c)$. Then Eqs.
(\ref{AA1}) and (\ref{AA2})  yield the parametric dependence of the
accelerated foil coordinate on time
\begin{equation}\label{asympt}
t=(\psi +w_{0}\psi ^{2}+2w_{0}^{2}\psi ^{3}/3)/\omega _{0},
\end{equation}
\begin{equation}
x=(w_{0}\psi ^{2}+2w_{0}^{2}\psi ^{3}/3)(c/\omega _{0}).
\end{equation}
In the limit  $t\ll 2\pi n_{0}l_{0}m_{p}c/E_{0}^{2}$ we have $x\approx
cw_{0}t^{2}/\omega _{0}$, and for  $t\rightarrow \infty $, $x\approx ct$,
while the momentum $p$ increases according  to Eq. (\ref{rad-press2}).

The function $w(\psi )$ given by Eq. (\ref{eq-w-psi}) can be
interpreted as the normalized energy of the portion of the laser pulse that
has interacted with the moving foil by time $t$. Its maximum value is
$w_{\max }={\cal E}_{las,in}/{\cal N}_{p}m_{p}c^{2}$, where
\begin{equation}
{\cal E}_{las,in}=\frac{E_{0}^{2}Sc\tau _{las,in}}{4\pi }
\end{equation}
is the laser pulse energy, ${\cal N}_{p}=n_{0}l_{0}S$ is the number of
protons in the region, with area equal to $S$, of the foil irradiated by the
laser pulse.

   From the solution of Eq. (\ref{eq-p}) given by Eq. (\ref{eq-p-W})
in terms of $w$ we obtain for the kinetic energy of a proton initially
at rest
\begin{equation}\label{energy}
{\cal E}_{p\mathrm{kin}} \equiv (m_{p}^{2}c^{2}+p^{2})^{1/2} c-
m_{p}c^2 = m_{p}c^{2}w^{2}/(w+1/2).
\end{equation}
In the limits $w\ll 1$ and $w\gg 1$, we have respectively
\begin{equation}
{\cal E}_{p\mathrm{kin}} \approx 2m_{p}c^{2}w^{2},
\quad {\rm and} \quad {\cal E}_{p\mathrm{kin}}\approx m_{p}c^{2}w.
\end{equation}
The upper limit to the proton kinetic energy and, correspondingly, to the
efficiency of the laser-to-proton energy transformation can be obtained from
Eq. (\ref{energy}) by setting $w=w_{\max} = {\cal E}_{las,in}/{\cal
N}_{p}m_{p}c^{2}$:
\begin{equation}
{\cal E}_{p\mathrm{kin},\max}=\frac{2{\cal E}_{las,in}}
{2{\cal E}_{las,in}+{\cal N}_{p}m_{p}c^{2}}\;
\frac{{\cal E}_{las,in}}
{{\cal N}_{p}}\,.  \label{eq-e-limit1}
\end{equation}
We see that within this model almost all the energy of the laser pulse is
transformed into the energy of the protons if ${\cal E}_{las,in}\gg
N_{p}m_{p}c^{2}/2$:
\begin{equation}
{\cal E}_{p\mathrm{kin},\max} \simeq \frac{{\cal E}_{las,in}}{{\cal
N}_{p}}\, \; \simeq \; \frac{E_0^2 c \tau_{las,in}}{4 \pi n_0 l_0}.
\label{eq-e-limit}
\end{equation}

\noindent
It is worth noting that, when the RPD mechanism takes place, the
dependence of the proton kinetic energy on the laser intensity and
duration turns out to be linear. Moreover, ${\cal E}_{p\mathrm{kin},\max}$
does not depend on the size of the illuminated area $S$: within this
simplified one-dimensional approximation, an increase of the focal spot
$S$ and the laser energy ${\cal E}_{las,in}$ that keep constant the ratio
\begin{equation}
\frac{{\cal E}_{las,in}}{S} = \frac{E_0^2 c \tau_{las,in}}{4 \pi}
\end{equation}
result in an increase of the number of accelerated protons
${\cal N}_{p}=n_0 l_0 S$ without perturbing the proton energy
spectrum.
The acceleration length $x_{acc}\approx
ct_{acc}$ and the acceleration time $t_{acc}$ can be estimated  using
Eq.(\ref{eq-e-limit}) and the $t^{1/3}$ asymptotic dependence of the
proton energy on
time in Eq.(\ref{asympt})  as
\begin{equation}
t_{acc}\approx \frac{2}{3}
\left(\frac{{\cal E}_{las,in}}{{\cal N}_{p}m_{p}c^{2}}\right)^2 \tau_{las,in}.
\end{equation}

\subsection{Computer Simulations}
\label{sec:simulation}

Within the framework of the simplified 1D approximation used above,
the protons formally have a monoenergetic spectrum. A number of
processes such as the transverse inhomogeneity of the amplitude of the
laser pulse, the electron stochastization due to ``vacuum heating''
\cite{vac} and the subsequent proton layer expansion under the action
of the Coulomb repelling force may result in the broadening of the
proton energy spectrum or in inefficient acceleration.

\begin{figure}
\begin{center}
\resizebox{0.7\textwidth}{!}{
\includegraphics{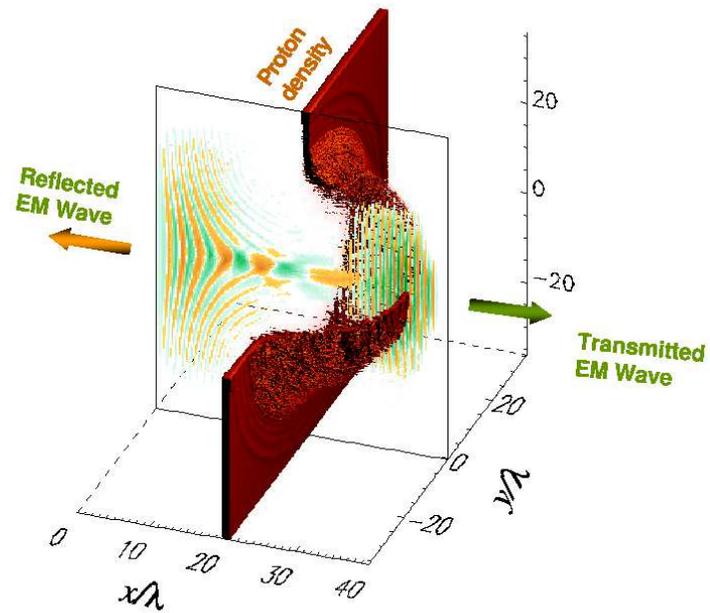} }
\end{center}
\caption{Spatial distribution of the proton density and the $z$
component of the electric field during the acceleration phase
(see text for details).}
\label{fig:protonden}
\end{figure}

\begin{figure}
\begin{center}
\resizebox{0.7\textwidth}{!}{
\includegraphics{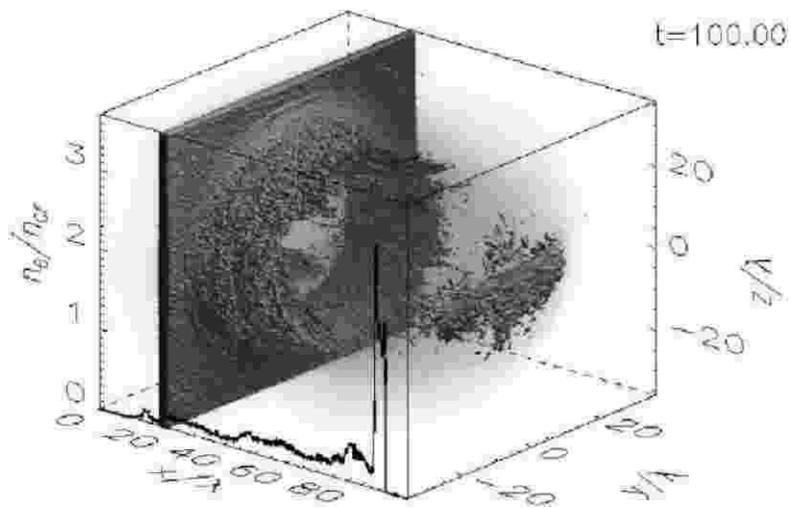} } 
\end{center}
\caption{Spatial distribution of the proton density at $t=
100$.}
\label{fig:acceleratedprot}
\end{figure}

\begin{figure}
\begin{center}
\resizebox{0.7\textwidth}{!}{
\includegraphics{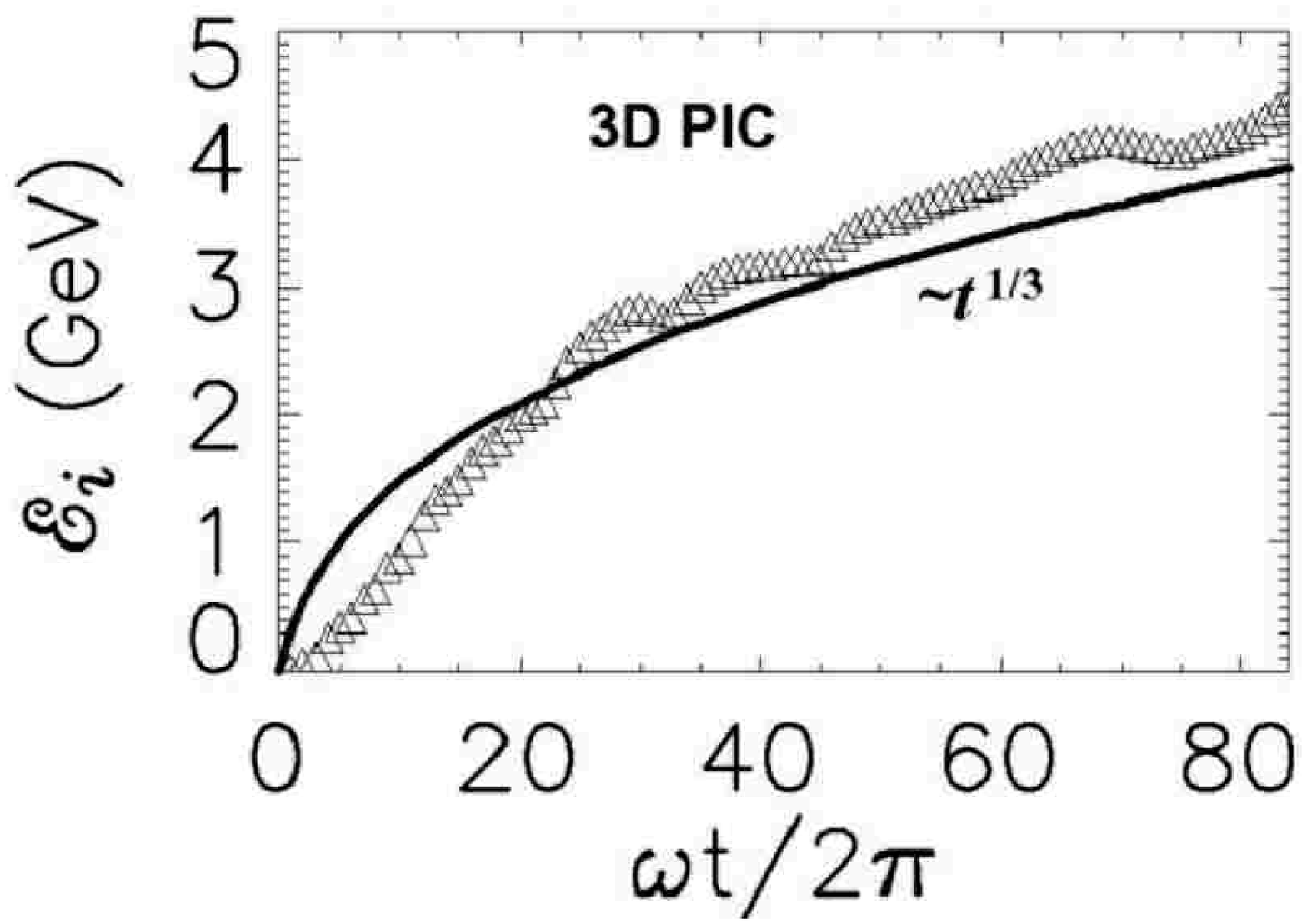} }
\end{center}
\caption{Maximum proton energy versus time. The continuous line
represents the analytical expectation.}
\label{fig:p_en_vs_time}
\end{figure}

\begin{figure}
\begin{center}
\resizebox{0.7\textwidth}{!}{
\includegraphics{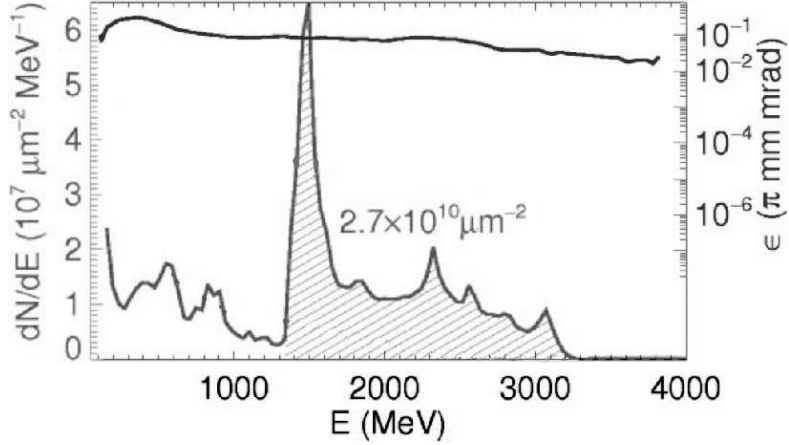} }
\end{center}
\caption{Proton energy distribution (red) and transverse emittance (blue) 
near the beam axis at $t=80$.}
\label{fig:p_axis}
\end{figure}

In order to examine this scheme in a three-dimensional geometry, whose
effects can indeed play a crucial role in the dynamics and stability
of the plasma layer under the action of a relativistically strong
laser pulse, we have performed\footnote{3D fully relativistic
simulations of laser-plasma interactions represent major numerical
efforts.  The simulation under consideration has been performed on 720
processors of the supercomputer HP Alpha server SC ES40 at JAERI
Kansai and form the basis of the studies of Refs.~\cite{Es}. Here we
make use of these results and, when appropriate, perform
extrapolations to other flux and energy ranges. A more detailed
numerical study is in progress and results will be presented in a
forthcoming publication.}  3D particle-in-cell (PIC) simulations with
the code REMP (Relativistic Electro-Magnetic Particle-mesh
code).  This code is based on the current assignment scheme ``Density
decomposition'' \cite{REMP}. In these simulations the laser pulse is
linearly polarized along the $z$-axis and propagates in the direction
of the $x$-axis. Its dimensionless amplitude is $a\equiv eE_0/m_e
\omega c=316$, corresponding to the peak intensity $I=1.37\times
10^{23}\, \mbox{W}/\mbox{cm}^{2}\times (1\,\mu \mbox{m} / \lambda 
)^{2}$, $\lambda=1 \mu\mathrm{m}$ being the laser wavelength. The
laser pulse is almost Gaussian with FWHM size $8\lambda \times
25\lambda \times 25\lambda $ and a sharp front starting from $a=100$;
its energy is ${\cal E}_{L}=10\,\mbox{kJ}\times (\lambda / 1\,\mu
\mbox{m})^{2}$.  Protons are generated from a $1 \ \mu m$
foil. In fact, due to the finite contrast of the laser, pulse
pedestals reaching the target before the main pulse will pre-form a
fully ionized plasma~\cite{ref:pedestalstudies,utsumi}.  Therefore, in
the present situation the target behaves as a fully ionized, $1\lambda
$ thick plasma with density $n_{e}=5.5\times
10^{22}\,\mbox{cm}^{-3}\times (1\,\mu \mbox{m}/\lambda )^{2} $, which
corresponds to the Langmuir frequency $\omega _{pe}=7\omega $. The
protons and the electrons have the same absolute charge and their mass
ratio is $m_{p}/m_{e}=1836$. The simulation box size is
$100\lambda\times 72\lambda \times 72\lambda $ corresponding to the
grid size $2500\times 1800\times 1800 $, so the mesh size is
$0.04\lambda $. The total number of quasi-particles is $4.37\times
10^{9}$. The boundary conditions are periodic along the $y$- and
$z$-axis and absorbing along the $x$-axis for both the e.m.  radiation
and the quasi-particles. The results of these simulations are shown in
Figs.~\ref{fig:protonden}-\ref{fig:p_axis}, where the space and time
units are respectively the wavelength $\lambda $ and period $2\pi
/\omega $ of the incident radiation.

Fig.~\ref{fig:protonden} shows the proton density and the $E_{z}$ 
component of the electric field. We see that a region of the foil with
the size of the laser focal spot is pushed forward. Although the
plasma in the foil is overcritical, it is initially ``transparent''
for the laser pulse due to the effect of relativistic transparency
(see e.g.  \cite{R1}). Therefore a portion of the laser pulse passes
through the foil. Eventually the pulse accelerates the electrons and,
as a result of the charge separation, a longitudinal electric field is
formed.  This can be interpreted as the ``rectification'' of the laser
light, by analogy with a rectifier in electrical engineering: the
transverse oscillating electric field in the pulse is transformed into
a longitudinal quasistatic electric field.  The dimensionless
amplitude of the longitudinal field is $a_{\parallel }\approx 150$
corresponding to $E_{||}=4.8\times 10^{14}\, \mbox{V}/\mbox{m}\times
(1\,\mu\mbox{m}/\lambda )$. The typical distance over which charge
separation occurs is comparable with the initial thickness of the foil
and is much smaller than the transverse size of the region that is
being pushed.  The proton layer is accelerated by this longitudinal
field. We note that the laser pulse frequency in the reference frame
comoving with the accelerated plasma region decreases as time
progresses so that the accelerating foil become less transparent with
time.

As seen in the cross-section of the electric field component $E_{z}$
in Fig.~\ref{fig:protonden}, the thickness of coloured stripes, which
corresponds to half of the radiation wave length, increases from left
to right in the reflected part of the pulse (along the $x$-axis). This
increase is weaker at the periphery (in the transverse
direction). This `nonuniform red shift' results from the Doppler
effect when the laser light is reflected from the co-propagating
relativistic mirror which accelerates and deforms in time. The red
shift testifies that the laser pulse does indeed lose its energy by
accelerating the plasma mirror. In this stage, the foil is transformed
into a ``cocoon'' inside which the laser pulse is almost confined. The
accelerated protons form a nearly flat ``plate'' at the front of the
``cocoon'' as is seen in Fig.~\ref{fig:acceleratedprot}.

Fig.~\ref{fig:p_en_vs_time} shows the maximum proton energy versus
time. This dependence is initially linear and, at later times, the
maximum proton energy scales as $t^{1/3}$ as predicted by the 1D
analytical model of Sec.~\ref{sec:1d}.  The protons in the plate
structure are accelerated according to the RPD regime.  These results
provide numerical evidence of the fact that the RPD acceleration
mechanism should appear already for laser intensities of
$10^{23}$~W/cm$^2$. The time evolution is hydrodinamically stable and
the acceleration highly efficient. In comparison with the experimental
results of present day Petawatt lasers, this example predicts yet
another astonishing advantage of the Exawatt lasers
\cite{tajima,Ross}, besides those described in
Refs.~\cite{tajima,MTB}.

In Fig.~\ref{fig:p_axis} we show the proton energy distribution
obtained in the 3D PIC simulations and their transverse emittance. The
energy spectra have been calculated for the particles in the region
near the beam axis (``plate'') within a $1 \mu m$ radius. We see that
the protons have a finite-width spectrum, localized within the
interval 1.3~GeV ${<} {\cal E}_{p\mathrm{kin}} {<}$ 3.2~GeV . The
number of protons, integrated over energy, in the dashed region in
Fig.  \ref{fig:p_axis} is equal to $2.7 \times 10^{10}$ particles per
$\mu m^2$. The proton transverse emittance is almost constant and
equals $\approx 10^{-2}\pi ~mm~mrad$.

The numerical studies depicted above are extremely challenging even
for large parallel computer facilities.  In order to reduce
complexity, the study has been carried out with laser pulse of
relatively small focal spot. In addition, the dynamical evolution has
been followed up until the $t^{1/3}$ asymptotic behaviour is reached
(i.e. before the complete laser-plasma decoupling).  The overall laser
energy to proton kinetic energy conversion efficiency ($\epsilon$) at
that time is 40\%. Extrapolation up to the time of decoupling
indicates that an energy conversion efficiency of 57\% can be
reached. On the other hand, due to the small focal spot, the RPD
regime is operational only the central area: here a
nearly-monochromatic spectrum is observed. The peripheral area
(``cocoon'') result in a nearly-thermal spectrum of protons.  Hence,
if we consider only the RPD accelerated protons at time of decoupling,
the efficiency drops to 11\%. This value is the most conservative
estimate of conversion efficiency and it is not expected to hold for
large focal spots $S$: the ``cocoon'' is the outcome of a border
effect due to the finite waist of the pulse and will not scale
linearly with $S$. Finally, note that the intensity and the duration
of the pulse has not been optimized for the production of protons in
the few GeV range but has been chosen to demonstrate the possibility
of highly relativistic ion generation~\cite{Es}. Awaiting for further
numerical studies, in the following sections the proton and neutrino
fluxes are provided as a function of the repetition rate and realistic
expectations are discussed below.

\subsection{Proton fluxes}

A proof of principle of the RPD acceleration mechanism is at the
borderline of current technology, but the possibility of using this
technique to overcome the limitation of traditional proton
accelerators faces many additional difficulties.  Present day systems
based on Chirped Pulse Amplification~\cite{CPA} (CPA) are able to
deliver intensities of the order of $10^{22}$ W/cm$^2$. It can be
expected that the intensity needed to trigger the RPD mechanism will
be reached in the near future since for intensities up to $10^{23}$
W/cm$^2$ the saturation fluence remains below the damage threshold and
we can still profit of CPA for short-pulse generation.  On the other
hand, all present high power lasers operate at very low repetition
rate. This is a classical problem e.g. in inertial fusion: here, the
basic principles could be demonstrated through the construction of
optical cavities delivering up to 1.8 MJ as the National Ignition
Facility (NIF) in US or the Laser Megajoule (LMJ) in France. However,
the ultimate use of ICF to produce electric power will require
repetition rates of the order of tens of Hz, an increase of several
order of magnitude compared to the shot rate achievable with
state-of-the-art fusion laser technology. This rate cannot achieved
with flashlamp-pumped neodymium-doped glass lasers that requires a
significant interpulse cooling time. There is, presently, a very large
effort to find alternative solutions~\cite{reprate}. We note here that
the solution of the problem of thermal stability of the system for
energy yield of tens of MJ/s would simultaneously provide an
appropriate driver for ICF operating with $\sim$ 2MJ laser pulse at a
repetition rate of ${\cal O}(10)$~Hz and, through the exploitation of
CPA and the RPD acceleration mechanism described above, an unsurpassed
proton acceleration facility for hadron and spallation neutron
production operating at energies of $\sim$1~GeV (laser pulses of
16~kJ/$\epsilon$) and repetition rates of the order of a few kHz.  As
a purpose of illustration, Tab.~\ref{proto_comp} compares the main
parameters of present and future proton drivers at the GeV energy
range. The LAMPF and ISIS beam dump facilities have given neutrino
sources for the LSND~\cite{Athanassopoulos:1996ds} and
KARMEN~\cite{Gemmeke:ix} experiments. Currently, two projects (JAERI
in Japan~\cite{JAERI_proposal} and SNS in USA~\cite{sns_proposal}) are
under construction, while the European Spallation Source project
(ESS)~\cite{ess_proposal} and the CERN Superconducting Proton Linac
(SPL)~\cite{SPL_yellowrep} are still pending approval.  A RPD
laser-driven acceleration facility (LAF) operating at 1~kHz and
providing $10^{14}$ proton/pulse (i.e. with a ``plate'' radius
of $\sim$34~$\mu m$) is also shown. It corresponds to a energy yield of
about 50~MJ/s at $\epsilon=0.3$. It is a remarkable fact that the RPD
acceleration mechanism allow a close synergy between the technological
needs for a ICF laser driver and an ultra-intense proton driver for
nuclear and particle physics applications.

\begin{table}
\caption{Main parameters of present and future proton drivers
at the GeV energy range. Pot~$\equiv$~proton on target.}
\begin{center}
\begin{tabular}{c|c|c|c|c|c|c}
\hline
         & Energy & pot/pulse      & Beam       & Repetition &Pulse
& Target       \\
         &  (GeV) & $\times10^{13}$& current    & rate       &width
&              \\
\hline
ISIS  & 0.8    & 2.5            & 200~$\mu$A & 50~Hz      &100~ns
& Tantalum     \\
LAMPF & 0.8    & 5.2            & 1~mA       &120~Hz      &
600~$\mu$s& Water/High Z \\
JAERI & 3.0    & 8.3             &333~$\mu$A  &
25~Hz      &1~$\mu$s   &  Hg          \\
SPL & 2.2    & 15             &1.8~mA  &
75~Hz      &2.2~ms   &  various          \\
SNS   & 1.0    & 15             & 1.4~mA     &
60~Hz      &695~ns     &  Hg          \\
ESS   & 1.3  & 84             &6.7~mA      &
50~Hz      &1.4~$\mu$s &  Hg          \\
LAF   & 1.0    & 10             &  16~mA     &  1 kHz     &$<$1~ps
& Water/High Z \\
\hline
\end{tabular}
\label{proto_comp}
\end{center}
\end{table}

\section{Neutrino beams from $\pi^+$ and $\mu^+$ decay-at-rest (DAR) and 
decay-in-flight (DIF)}
\label{sec:nubeams}

\subsection{The  $\pi^+$ and $\mu^+$ decay chains}
In a neutrino beam-line based on a dump of low energy protons into a
passive material, neutrinos arise from both pion and muon decays. The
production of kaons or heavier mesons is negligible if the proton
energy is sufficiently low ($E_p\simle 3$~GeV). Therefore, the neutrino
beam does not suffer from $\nu_e$ and $\bar{\nu}_e$ contaminations due
to kaon decays.  The pion decay modes are
$\pi^+\rightarrow\mu^+\nu_\mu$, $\pi^+\rightarrow e^+\nu_e$,
$\pi^-\rightarrow\mu^-\bar{\nu}_\mu$ and $\pi^-\rightarrow
e^-\bar{\nu}_e$ but the decays into electrons are strongly suppressed.
The muon decay modes are $\mu^+\rightarrow e^+\nu_e\bar{\nu}_\mu$ and
$\mu^-\rightarrow e^-\bar{\nu}_e{\nu}_\mu$. Almost all $\mu^+$ stop
before decaying and produce a Michel spectrum for $\nu_e$ and
$\bar{\nu}_\mu$ while the $\mu^-$ are captured in orbit.  The $\pi^+$
decay occurs both with the pion at rest, providing a mono-energetic
neutrino spectrum, and in flight. The ratio DAR/DIF depends on the
material and the geometry of the target. Proton-rich targets
(e.g. water) are employed to obtain large pion yields. Early stopping
of the mesons is achieved positioning a dense dump just after the
water vessel. The distance between the water target and the stopper
can be tuned to optimize the DAR/DIF ratio.

\subsection{Neutrino energy spectra and beam composition}
\label{neutrino_e_spectra}

In order to estimate the expected neutrino energy spectra and beam
composition, we refer to the setup of the LSND experiment, operated at
LAMPF from 1993 to 1998 with 800~MeV protons impinging into a water
target followed by a copper beam stopper.  The precise evaluation for
a laser driven facility should include the secondary yield of the
nearly monochromatic distribution plus the higher energy tail
(Fig.~\ref{fig:p_axis}) and the contribution of the quasi-thermal
spectrum.  The former will be peaked in the 1-2~GeV range to make
fully operative the RPD mechanism. The size of the latter will depend
on the cocoon/plate ratio, as described in Sec.~\ref{sec:simulation}.
Clearly, a precise determination of the fluxes is beyond the scope of
this paper; however, we note that the use of LAMPF data implies a
significant underestimation ($\sim 50\%$) of the $\pi^+$ yield and a
small underestimation of the background from DIF which can be reduced
through a dedicated optimization of the target-stopper distance.  In
the following we refer to a laser-driven facility providing $10^{14}$
protons-on-target (pot) per pulse at different repetition rates with a
nearly monochromatic spectrum corresponding to the LSND
setup\footnote{Full flux calculations are available
in~\cite{sito_LSND}.  In fact, all proposed future neutron spallation
sources have as target material liquid mercury.  The reason for such a
choice is twofold: mercury is liquid at room temperature. Therefore,
its recirculation allows a more efficient power dissipation with
respect to solid target. Moreover, the high atomic number gives a
source of numerous neutrons.  For neutrino applications only the first
motivation holds. Indeed, high-Z elements are less efficient in
producing neutrinos that low-Z elements. On the other hands, high-Z
elements allow a strong suppression of DIF neutrinos. In water-based
targets this suppression, if needed, can be partially recovered
modifying the distance between the water vessel and the stopper.}.

In the LSND target configuration the DAR/DIF ratio turned out to be
97\%/3\% \cite{Athanassopoulos:1996wc,Athanassopoulos:1997er,Aguilar:2001ty}.
The leading decay chain $\pi^+ \rightarrow \mu^+ \nu_\mu \rightarrow
e^+ \nu_e \bar{\nu}_\mu \nu_\mu $ does not contain $\bar{\nu}_e$ and
offers a unique opportunity to test the occurrence of $\bar{\nu}_\mu
\rightarrow \bar{\nu}_e$ transitions.  The decay of $\pi^-$ might
lead, in principle, to a large $\bar{\nu}_e$ contamination. However,
three factors contribute to its suppression. At these proton energies,
the $\pi^+$ production rate is larger than $\pi^-$ by about a factor
8. Moreover, negative pions which come to rest are captured before
they decay: in fact, at LAMPF, only 5\% decay in orbit and, hence,
contribute to the $\bar{\nu}_e$ background. Finally, almost all
negative muons arising from the decays in flight come to rest in the
beam dump before decaying; most then undergo the reaction $\mu^-
N\rightarrow\nu_\mu e^-$ that leads to $\nu_\mu$ with energy below
90~MeV, leaving only 12\% of them to decay into $\bar{\nu}_e$.  Hence,
the relative $\bar{\nu}_e$ yield of the $\pi^-$ decays at LSND,
compared to the $\pi^+$ decays, is
$\sim(1/8)\times0.05\times0.12\approx7.5\times10^{-4}$.  Clearly, the
level of $\bar{\nu}_e$ contamination is much smaller than the
intrinsic $\nu_e$ contamination (a few \%) of a high energy $\nu_\mu$
beam from $\pi$ decay (e.g. the Superbeams).

The $\pi^+$ decay chain provides an intense source of mo\-no\-chro\-ma\-tic
muon neutrinos ($E=29.8$~MeV). Due to the short $\pi^+$ lifetime
($\tau_\pi \simeq 26$~ns) these neutrinos closely follow the beam time
profile. This fact opens up the possibility to detect $\nu_\mu \
\rightarrow \nu_e $ oscillations: the $\nu_\mu \ \rightarrow \nu_e $
oscillations can be temporally separated from the events due to
$\mu^+$ DAR electron neutrinos, which appear on a time scale of few
$\mu$s due to the muon lifetime ($\tau_\mu \simeq 2.2 \ \mu$s). In
order to exploit the well defined time structure of a DAR beam, it is
mandatory to have a proton pulse width comparable with the pion
life-time.  For a laser-driven facility this requirement is easily
fulfilled, the proton pulse temporal spread being of the order of 1~ps.

\begin{figure}
\begin{center}
\resizebox{0.7\textwidth}{!}{ 
\includegraphics{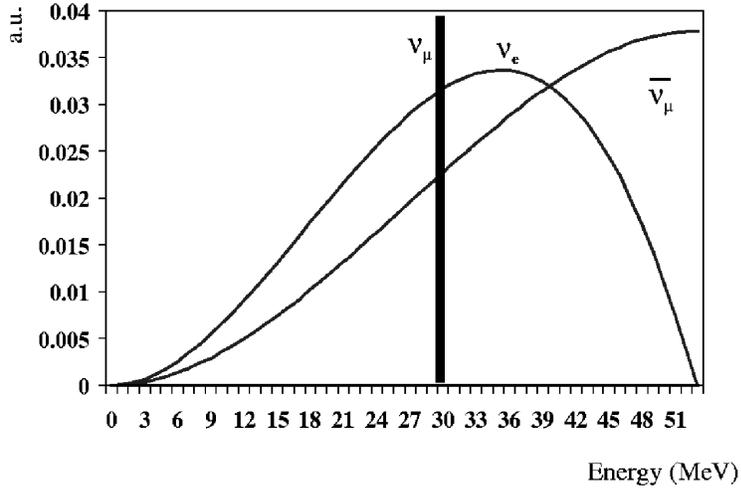} }
\end{center}
\caption{Neutrino energy spectra from $\pi^+ \rightarrow \mu^+ \nu_\mu
\rightarrow e^+ \nu_\mu \nu_e \bar{\nu}_\mu$ decay at
rest.} \label{darfluxes}
\end{figure}

Finally, the shape of the neutrino flux from $\pi^+$ and $\mu^+$ decay
at rest (DAR) is well known and it is shown in
Fig.~\ref{darfluxes}. Therefore, only the absolute amplitude has to be
determined from experiments and simulation.

The simulation of the expected neutrino fluxes, both at LAMPF and
ISIS, was performed by using the pion yield in proton-target
interaction measured in a dedicated
experiment~\cite{Allen:1989dt}. For details on the neutrino flux
simulation and the associated uncertainties we refer to
~\cite{Burman:1989dq,Burman:gt}. The main source of systematic error
is associated with the pion yield in proton-target interaction and it
has been estimated to be about 6\%. This has to be compared with the
total systematic error associated to the neutrino flux from decay at
rest that has been estimated to be of about
7\%~\cite{Burman:1989dq}. For neutrinos from decay in flight the
systematic error has been estimated to be about
15\%~\cite{Athanassopoulos:1997er}.

\section{The neutrino oscillation channels and the detector}
\label{sec:channel_detector}

In principle neutrinos from DAR and DIF decays could allow
simultaneous investigation of
\begin{itemize}
\item $\bar{\nu}_\mu \ \rightarrow \bar{\nu}_e$ oscillations through
the identification of DAR $\bar{\nu}_\mu$ transitions (``DAR analysis'');
\item $\nu_\mu \ \rightarrow \nu_e$ oscillations from high energy DIF
$\nu_\mu$ (``DIF analysis'');
\item  $\nu_\mu \ \rightarrow \nu_e$ oscillations from DAR
$\nu_\mu$ temporally separated with respect to DAR $\nu_e$ (``time analysis'').
\end{itemize}
So far these channels have been used to explore neutrino oscillations
at a $\Delta m^2$ of about 1~eV$^2$ ($E \sim 10^1$~MeV and $L \sim
10^1$~m). The LSND experiment gave a positive result (there is a claim
for an excess of events with an electron in the final state induced by
$\bar{\nu}_\mu$ ($\nu_\mu$) oscillations into $\bar{\nu}_e$
($\nu_e$)~\cite{Athanassopoulos:1996wc,Athanassopoulos:1997er,Aguilar:2001ty}),
while the KARMEN experiment gave a negative
result~\cite{Armbruster:2002mp}. The need for a check of these
experiments caused the proposal of new projects like
MiniBooNE~\cite{Stancu:14dr} (currently data taking) and of new
experiments at the Neutron Spallation Source~\cite{workshopnss}.
However, there are no proposals to search with DAR and/or DIF
neutrinos for oscillations at a baseline of few kilometers to test
sub-dominant $\nu_e$ and $\bar{\nu}_e$ appearance modes at the
atmospheric scale. This test is unfeasible with present
accelerators. Much higher intensities ($\mathcal{O}$($20 \, \rm{mA}$)
or more) are needed to overcome the large suppression due to the
smallness of the $\nu_e$~CC cross-section at these energies. As noted
before, a laser driven facility could offer this opportunity.

The optimization of the detector coupled with this facility is beyond
the scope of this paper.  Here, we will consider a detector with a
technology similar to LSND and with a fiducial mass comparable with
Super-Kamiokande.  This corresponds to $1.3\times 10^{33}$ free
protons (17~kton of CH$_2$). 
Similar detectors based on liquid scintillator have been recently
proposed for low-energy neutrino astronomy and proton decay~\cite{LENA}.
In the following, when appropriate, we
refer to the LSND experimental and Monte Carlo studies to assess the
physics performance of the
apparatus~\cite{Athanassopoulos:1996wc,Athanassopoulos:1997er,Aguilar:2001ty}.

The LSND detector consisted of a cylindrical tank filled with liquid
scintillator. 
The composition was chosen to be sensitive
to both Cerenkov light from electrons and relativistic muons and
scintillation light from all charged particles. The light was detected
through photomultiplier tubes (PMT's) covering 25\% of the
detector surface. 
PMT time and pulse-height
signals were used to reconstruct the track.
The Cerenkov cone for relativistic particles and
the time distribution of the light, which is broader for non
relativistic particles, gave excellent separation between electrons
and particles below Cerenkov threshold.

\section{The $\bar{\nu}_\mu\rightarrow\bar{\nu}_e$ decay at rest analysis}
\label{sec:dar_analysis}

\begin{figure}
\begin{center}
\resizebox{0.7\textwidth}{!}{
\includegraphics{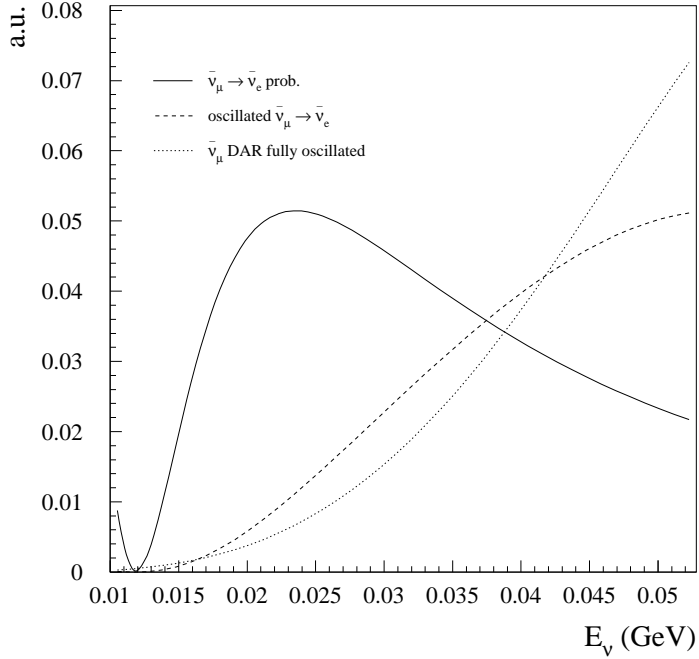} }
\end{center}
\caption{Energy spectrum of the fully oscillated
$\bar{\nu}_\mu\rightarrow\bar{\nu}_e$ (dotted line) before the
selection cuts.  The two-family oscillation probability for
$\Delta m^2=2.5 \times 10^{-3} \, \rm{eV^2}$ at L = 11 km
(continuous line) and the corresponding convoluted
$\bar{\nu}_\mu\rightarrow\bar{\nu}_e$ spectrum (dashed line)
are also shown.}
\label{fig:spectrum}
\end{figure}

The search for $\bar{\nu}_\mu\rightarrow\bar{\nu}_e$ oscillations is
performed by using $\bar{\nu}_\mu$ from $\mu^+$ decay at rest.
$\bar{\nu}_e$ are detected through the reaction $\bar{\nu}_e p
\rightarrow e^+ n$ (plus a small contamination of $\bar{\nu}_e C
\rightarrow e^+ \ B \ n$) followed by the neutron capture reaction $n
p\rightarrow d\gamma$, the $\gamma$ energy being 2.2~MeV.  A candidate
$\bar{\nu}_e$ events consists in one identified electron with energy
in the $20<E_e<60$~MeV range\footnote{$\beta$ decays of cosmogenic
$^{12}B$ prevent the use of the candidate with $E<20$~MeV.}  and one
associated gamma.  The electron identification efficiency at LSND is
42\% with a relative systematic error of 7\%. Note that the $e^+$
inefficiency is dominated by the need of vetoing Michel electrons from
cosmic muons; in particular the dead-time of the veto accounts for an
electron efficiency reduction of 24\%~\cite{Aguilar:2001ty}. The size
of this background strongly depends on the shallow depth of the
experimental area where LSND is located and would be significantly
suppressed at deeper locations. The correlated photon is identified
combining the information from the number of PMT's hits associated
with the $\gamma$, its distance from the positron and the time
interval between the $e^+$ and the $\gamma$ which exploits the $186\
\mu$s delay for the neutron capture in mineral oil. The LSND analysis
is based on a likelihood function $R_\gamma$ whose $R_\gamma > 10$ cut
corresponds to an efficiency of 39\% and a contamination due to
accidentals at the level of 0.26\%. The corresponding relative
systematic uncertainty does not exceed 7\%.  Fig.~\ref{fig:spectrum}
shows the energy spectrum of $\bar{\nu}_\mu\rightarrow\bar{\nu}_e$
oscillated neutrinos for 100\% conversion probability before any
selection cut:

\be
\Phi_{\bar{\nu}_\mu}(E) \ \sigma_{\bar{\nu}_e  p \rightarrow e^+ n}(E)
\ee
\noindent i.e. the product of the $\bar{\nu}_\mu(E)$ flux from $\mu^+$
DAR and the $\bar{\nu}_e p \rightarrow e^+ n$ CC cross-section.  The
two-family oscillation probability for $\Delta m^2=2.5 \times 10^{-3}
\, \rm{eV^2}$ at L = 11 km and the corresponding

\be
\Phi_{\bar{\nu_\mu}}(E) \ \sigma_{\bar{\nu}_e  p \rightarrow e^+ n}(E)
P(\bar{\nu}_\mu \rightarrow \bar{\nu}_e)(E)
\ee

\noindent product is also shown. The corresponding cross-section,
weighted with the DAR spectrum, is $0.95\times10^{-40}~\mbox{cm}^2$.

  Contaminations to the
$\bar{\nu}_e$ sample arise from beam-related interactions with
neutrons in the final state, beam events without neutrons and
background produced by cosmics and radioactivity (``beam
unrelated'').

The main beam-related backgrounds with final state neutrons is
$\bar{\nu}_e \ p \rightarrow e^+ n$ from $\mu^-$ decays at rest. As
noted before, the $\mu^-$ DAR yield is suppressed by a factor
$7.5\times10^{-4}$ w.r.t.  $\mu^+$ DAR. The $\bar{\nu}_e \ p
\rightarrow e^+ n$ cross-section weighted according to the
$\bar{\nu}_e$ DAR spectrum is
0.$72\times10^{-40}$~cm$^2$~\cite{ref:llewellyn}. The fraction of
events with energy greater than 20~MeV is 0.806 while the electron and
$\gamma$ efficiencies are the same as for the signal sample.  The
second most important source of beam-related background events with
correlated neutrons is the misidentification as $\bar{\nu}_e$ events
of $\bar{\nu}_\mu \ p \rightarrow \mu^+ n$ ~CC interactions from
$\pi^-$ DIF. Because of the energy needed to produce a $\mu^+$, such a
$\bar{\nu}_\mu$ must arise from a decay in flight. The final state
considered are $\bar{\nu}_\mu p \rightarrow \mu^+ n$ or (less often)
$\bar{\nu}_\mu C \rightarrow \mu^+ n X$, followed by $\mu^+
\rightarrow e^+ \nu_e \bar{\nu}_\mu $.  In most of the cases, the muon
is missed because it decays at very large times compared with
$\tau_\mu \simeq 2.2 \ \mu\rm{s}$ or the deposited energy is below the
phototube threshold.  The latter can occur either because the muon is
too low in energy or it is produced behind the phototube surfaces. For
this background, the flux weighted cross-section is $4.9\times
10^{-40}$~cm$^2$, the fraction of muons in the tail of the lifetime
distribution ($\tau>12 \ \mu$s or with very low kinetic energy ($T<3$
MeV)) is 2.6\%. The positron efficiency is 42\%, the fraction of
events with $E>20\,{\rm MeV}$ is 81.6\% and, again, the efficiency for
correlated $\gamma$ is 39\%.  Other source of misidentification are
muon decays in the tail of the lifetime distribution, prompt decays to
electrons so that the $\mu$ and the $e$ are collected in a single
event and muon lost by trigger inefficiencies. The whole background
from $\mu$ misidentification has been computed according to the
results of~\cite{Aguilar:2001ty}.  However, the DIF flux has been
corrected keeping into account the larger distance of the detector (L
= 11~km versus 30~m): at large baselines the DIF fluxes are similar to
the corresponding fluxes at the center of LSND corrected for the
$L^{-2}$ suppression term.

The main source of beam related background without correlated neutrons
is $\nu_e ~^{12}C \rightarrow e^- X$ scattering. The corresponding
average cross-section is $1.5 \times 10^{-41}$
cm$^2$~\cite{Kolbe:xb}. For an electron reconstruction efficiency of
0.36~\cite{Athanassopoulos:1996wc}, the fraction of events with $E>20$
MeV is 46\% while the electron efficiency and the accidental $\gamma$
efficiency at LSND are, respectively, 42\% and 0.26\%. Other
sub-dominant sources are discussed in~\cite{Athanassopoulos:1996wc}.

Beam unrelated background results mainly from unvetoed cosmic
interactions in delayed coincidence with accidental photons. It
strongly depends on the veto quality, the depth of the detector
and the rate of accidental photons. Moreover, it can be suppressed
if the duty cycle of the beam is sufficiently low. For the case of
LSND, the poor time structure of the beam and the shallow 
depth of the detector does not allow an effective
suppression of this background, which is estimated during the
beam-off data taking and, hence, subtracted.  On the other hand,
the time structure of ISIS allows a suppression of more than two
order of magnitude compared with LSND in the
$\bar{\nu}_\mu\rightarrow\bar{\nu}_e$ channel, so that KARMEN does
not suffer from this contamination. For the laser-driven
facilities considered here the time structure is very well defined
(see Sec.~\ref{sec:laser} and \cite{pakhomov}) and the beam-off
background could be non negligible (at the depth of LSND) only for
very high repetition rates ($\sim 10 \, \rm{kHz}$). At larger
depths, this contamination is negligible for any realistic
repetition rate and, hence, it is not considered in the present
analysis.

\section{Sensitivity to $\theta_{13}$}
\label{sec:theta_13}

In order to get information about the magnitude and phase of the
$U_{e3}$ term of the PMNS mixing matrix, terrestrial experiments
explore sub-dominant effects in the neutrino transition probabilities
at the atmospheric scale which, in general, are suppressed by at least
one power of $\alpha \equiv \Delta
m^2_{21}/ | \Delta m^2_{23} |$ \footnote{A global analysis of all
available neutrino oscillation data gives for $\alpha$ a best fit
value of 0.026, while the 3$\sigma$ allowed range is $0.018 < \alpha <
0.053$~\cite{Maltoni:2003da}.}. Since matter effects are negligible
for baselines of the order of $L = 10$~km, the $\nu_\mu \rightarrow
\nu_e$ oscillation probability can be expressed as~\cite{cervera_freund}:
\begin{eqnarray}
\label{eq:PROBVACUUM}
P(\nu_\mu \rightarrow \nu_e) 
 \simeq  \sin^2 2\theta_{13} \, \sin^2 \theta_{23} 
\sin^2 {\Delta} & &  \nonumber \\
-  \alpha\; \sin 2\theta_{13} \, \sin\delta  \, \cos\theta_{13} \sin
2\theta_{12} \sin 2\theta_{23}
\sin^3{\Delta} & & \nonumber \\
-  \alpha\; \sin 2\theta_{13}  \, \cos\delta \, \cos\theta_{13} \sin
2\theta_{12} \sin 2\theta_{23}
    \cos {\Delta} \sin^2 {\Delta} & & \nonumber  \\
+ \alpha^2 \, \cos^2 \theta_{23} \sin^2 2\theta_{12} \sin^2 {\Delta} 
& & \nonumber \\
 \equiv  O_1 + O_2 + O_3 + O_4, & &
\end{eqnarray}
$\Delta$ being the oscillation phase $\Delta m^2_{23} \ L/4E_\nu $ in
natural units ($c=\hbar=1$).  If the energy and the baseline of the
experiment is chosen to fulfill $\Delta \simeq \pi/2$ (oscillation
maximum), the $O_3$ term is suppressed. Similarly, the term $O_4$ can
be neglected unless $\sin^2 2 \theta_{13}\simeq$
$\mathcal{O}$($10^{-4}$).  Hence, the $\nu_\mu \rightarrow \nu_e$
transition probability can be expressed in the simplified form:
\begin{eqnarray}
\label{eq:PROBVACUUM2}
P(\nu_\mu \rightarrow \nu_e)
& \simeq & \sin^2 2\theta_{13} \, \sin^2 \theta_{23}
\sin^2 {\Delta} \nonumber \\
&-&  \alpha\; \sin 2\theta_{13} \, \sin\delta  \, \cos\theta_{13} \sin
2\theta_{12} \sin 2\theta_{23}
\sin^3{\Delta}
\end{eqnarray}
Its CP conjugate  $\bar{\nu}_\mu \rightarrow \bar{\nu}_e$ is therefore
\begin{eqnarray}
\label{eq:PROBVACUUM3}
P(\bar{\nu}_\mu \rightarrow \bar{\nu}_e)
& \simeq & \sin^2 2\theta_{13} \, \sin^2 \theta_{23}
\sin^2 {\Delta} \nonumber \\
&+&  \alpha\; \sin 2\theta_{13} \, \sin\delta  \, \cos\theta_{13} \sin
2\theta_{12} \sin 2\theta_{23}
\sin^3{\Delta}
\end{eqnarray}
and a simultaneous measurement of $\nu_e$ and $\bar{\nu}_e$ gives
access\footnote{This determination is not unique due to the $\delta
\rightarrow \pi - \delta$ and $\theta_{23} \rightarrow \pi/2-
\theta_{23}$ invariance of the transition
probabilities~\cite{burguet-castell,barger_theta_12_amb}.} to the
magnitude and phase of the $U_{e3}$ entry.

\begin{figure}
\begin{center}
\resizebox{0.7\textwidth}{!}{
\includegraphics{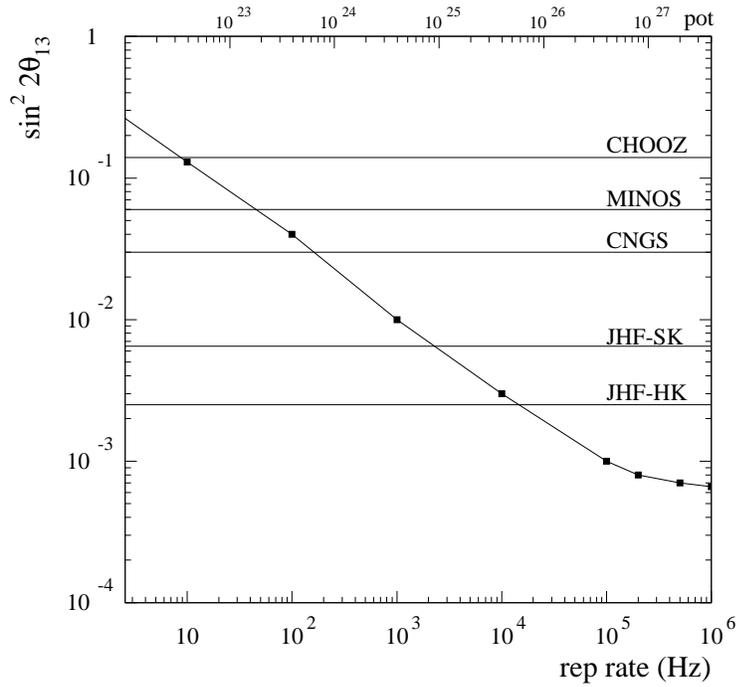} }
\end{center}
\caption{Exclusion limit at 90\% CL for $\sin^2 2\theta_{13}$
(two parameter fit) as a function of the repetition rate. The
corresponding integrated protons on target for 5y of data
taking are also shown. The horizontal bands indicate the
corresponding limit from $\nu_\mu \rightarrow
\nu_e$ appearance search at MINOS, CNGS,
JPARC to Super-Kamiokande and JPARC to Hyper-Kamiokande;
the upper band shows the present CHOOZ limit from $\bar{\nu}_e$
disappearance search.}
\label{fig:sens}
\end{figure}

\begin{figure}
\begin{center}
\resizebox{0.7\textwidth}{!}{
\includegraphics{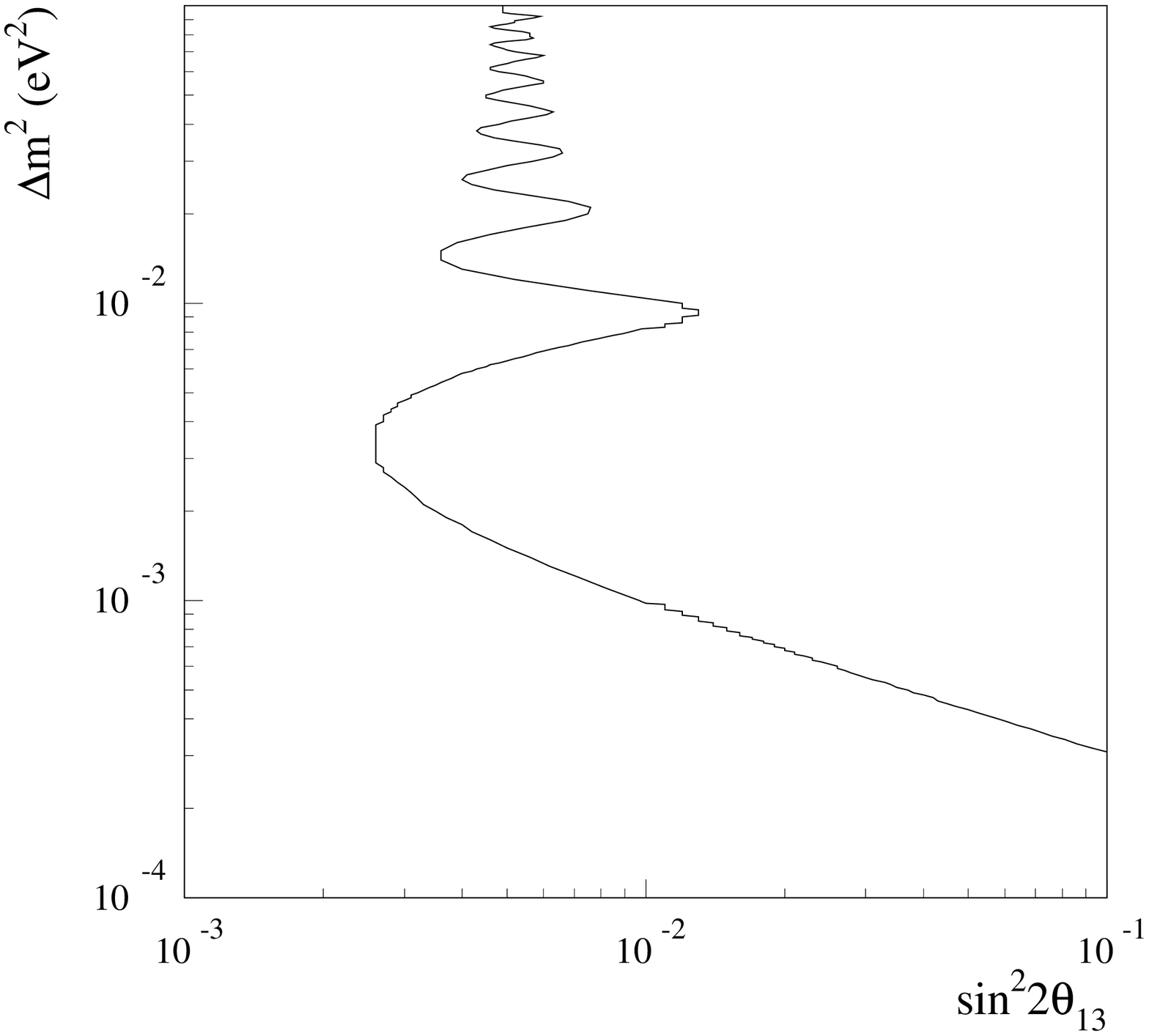} }
\end{center}
\caption{Exclusion limit at 90\% CL in the
$(\Delta m^2_{23},\sin^2 2\theta_{13})$ plane  for 10 kHz
repetition rate.}
\label{fig:excluded}
\end{figure}

The analysis described in Section~\ref{sec:dar_analysis} allows a
determination of $P(\bar{\nu}_\mu \rightarrow \bar{\nu}_e)$. To ease
comparison with other proposed facilities, we interpret
$P(\bar{\nu}_\mu \rightarrow \bar{\nu}_e)$ assuming no CP violation in
the leptonic sector ($\delta = 0$), $\theta_{23}=\pi/4$ and $\Delta
m^2_{23}=2.5 \times 10^{-3} \, \rm{eV^2}$, so that

\begin{eqnarray} 
P(\bar{\nu}_\mu \rightarrow \bar{\nu}_e)  \simeq  \sin^2
2\theta_{13} \sin^2 \theta_{23} \sin^2 \Delta \nonumber \\  =  \frac {1}{2}
\sin^2 2\theta_{13} \sin^2 \left[1.27 \frac{\Delta
m^2_{23}(\rm{eV}^2) \ L(\rm{km})}{E(\rm{GeV})} \right] 
\end{eqnarray}

\noindent and performing a two parameter fit of $\Delta m^2_{23}$ and
$\sin^2 2\theta_{13}$. The fluxes of $\bar{\nu}_\mu$ from $\pi^+$ DAR
are computed for a facility providing $10^{14}$ pot/pulse with
variable repetition rates. The $\pi^+$ yield considered here is
0.09~$\pi^+$/pot, corresponding to the LSND setup
~\cite{Burman:1989dq}. As discussed in Sec.~\ref{neutrino_e_spectra},
this estimate is rather conservative for a laser-based facility. In 5
years of data taking, assuming 6 months of operation and 50\% beam on
time, the overall integrated flux at 11~km is $2.3\times10^7 \times R\
\nu/\mbox{cm}^2$, $R$ being the repetition rate in Hz.  At $R\ =
1$~kHz, we expect 473 $\bar{\nu}_e$~CC events assuming 100\% $
\bar{\nu}_\mu \rightarrow \bar{\nu}_e$ conversion rate.  The
corresponding background is below 0.6 events (0.28 from $\mu^-$ DAR,
0.16 from $\pi^-$ DIF with the muon misidentified and 0.16 from events
without correlated $\gamma$). The systematic errors for signal and
background are inferred from the LSND analysis and shown in
Table~\ref{tab:sys_dar}. Fig.~\ref{fig:sens} shows the 90\% confidence
level (CL) exclusion limit for $\sin^2 2\theta_{13}$ in the occurrence
of the null hypothesis ($\theta_{13}=0$) as a function of the
repetition rate. The corresponding sensitivities coming from the
$\nu_\mu \rightarrow \nu_e$ appearance search for MINOS~\cite{minos},
CNGS~\cite{Komatsu:2002sz}, JPARC-SK~\cite{JHF} and JPARC to
Hyper-Kamiokande~\cite{JHF} are also shown, together with the present
CHOOZ~\cite{CHOOZ} limit from $\bar{\nu}_e$ disappearance.  The
two-parameter exclusion region at 90\% CL for a 10~kHz facility is
depicted in Fig.~\ref{fig:excluded}.

\begin{table}
\caption{Breakdown of systematic errors for the $\bar{\nu_\mu} \rightarrow
\bar{\nu_e}$ DAR analysis.}
\label{tab:sys_dar}
\begin{center}
\begin{tabular}{l|c|c}
\hline
Channel & Source       & Error (\%) \\ \hline
signal  & flux         & 7          \\
           & cross-sec    & -          \\
           & eff          & 7          \\
           & $\gamma$ id  & 7          \\  \hline
$\bar{\nu}_e$ from DAR $\mu^-$ & flux         & 15 \\
                                  & cross-sec    & -  \\
                                  & eff          & 7  \\
                                  & $\gamma$ id  & 7  \\ 
\hline
$\bar{\nu}_\mu$ from DIF $\pi^-$ & flux         & 15 \\
                                    & eff + cross-sec & 41 \\
                                  & $\gamma$ id  & 7  \\ 
\hline
$\nu_e \ ^{12}C$ from DAR $\mu^+$  & inclusive    & 17 \\
\hline\noalign{\smallskip}
\end{tabular}
\end{center}
\end{table}

\section{$\nu_\mu \rightarrow \nu_e$ appearance searches}

In LSND $\nu_\mu \rightarrow \nu_e$ oscillations were searched for by
exploiting the $\nu_\mu$ flux from decay in flight. As discussed in
Section~\ref{sec:nubeams}, about 3\% of the $\pi^+$ decay in flight
originating a $\nu_\mu$ beam with average energy of about
90~MeV. $\nu_e$ are detected through the reaction $\nu_e\ e\rightarrow
e\ X$ and requiring the electron to have energy in the range
$60\div200$~MeV. The corresponding value of the cross-section for
$<E_\nu>\sim 9$~MeV is about
$15\times10^{-40}~\mbox{cm}^2$~\cite{Kolbe:xb}, i.e. a factor 15
larger than $\bar{\nu}_e$ inverse $\beta$-decay (see
Section~\ref{sec:dar_analysis}). Taking into account the different
detection efficiencies~\cite{Athanassopoulos:1997er} we expect a ratio
of fully oscillated DIF/DAR events of about one third with a signal to
noise ratio comparable to both analyzes. Note, however, that the
energy of DIF neutrinos does not match the maximum of the oscillation
probability for the baseline considered ($L = 11$~km for $\Delta
m^2_{23} = 2.5\times 10^{-3}~\mbox{eV}^2$).

The time analysis searches for $\nu_\mu \rightarrow \nu_e$
oscillations by looking for mono-energetic $\nu_e$ ($E_\nu =
29.8$~MeV) in a short time window after the proton pulse. The length
of the time window is determined by the pion life-time ($\tau_\pi =
26$~ns). The $\nu_e$ are detected through the CC reaction of $\nu_e$
onto $^{12}C$ giving rise to an electron and a $^{12}N$ (ground state)
followed by the $\beta$-decay $^{12}N_{g.s.}\rightarrow\ ^{12}C\ e^+\
\nu_e$ with 15.9~ms decay time.  This analysis was found particularly
appealing for the following reasons: the main background is induced by
$\nu_e$ from fast $\mu^+$-decay within the time window, but can be
precisely measured and subtracted; due to $\nu_e \ ^{12}C \rightarrow
e^- \ ^{12}N_{g.s.}$ interactions from $\mu^+$ decays outside the time
window, the full oscillation expectation is normalized by the
experiment itself. However, the sensitivity of this channel is
essentially limited by the small cross-section of the involved
reaction ($4.95\times10^{-42}~\mbox{cm}^2$, at $E_\nu = 29.8$~MeV, to
be compared with the inverse $\beta$-decay reaction whose
cross-section is $0.95\times10^{-40}~\mbox{cm}^2$, convoluted over the
whole DAR spectrum).
Summarizing the search for $\nu_\mu \rightarrow \nu_e$ oscillations is
very interesting for CP-violation studies, but it is not at the peak
of the oscillation probability in the DIF analysis
(E/L $\sim$ 0.1~GeV/11~km $\sim10^{-2}$; $\Delta\sim 0.34$) and it is
limited by statistics in the time analysis. The impact of this channel
on CP-violation studies is currently under investigation and will be
the subject of a forthcoming paper.

\section{Conclusions}
The current debate for multipurpose facilities aimed at high precision
studies of neutrino oscillations drove the authors towards the study of
non-conventional neutrino sources.  In particular, we noted that the
possibility to accelerate efficiently protons in the GeV energy range
thro\-ugh relativistic laser-plasma interactions opens up interesting
opportunities for the development of a new generation of proton
drivers.  In this paper we discussed a radiation pressure dominated
(RPD) mechanism for relativistic proton acceleration. This mechanism
is highly efficient compared to previous proposals and could allow a
close synergy between present R\&D finalised to energy production
through inertial confined fusion and the wealth of applications
related to high intensity multi-GeV drivers (see Appendix). Moreover,
we demonstrated that this facility could allow for the first time the
study of subdominant $\nu_\mu \rightarrow \nu_e$ oscillations at the
atmospheric scale with neutrinos produced by $\pi^+$ decays at rest or
in flight.

\section*{Acknowledgements}
We wish to thank S.~Atzeni, M.~Borghesi, A.~Donini, J. Koga, K. Nishihara,
A. V. Titov, M. Yamagiwa for many useful suggestions
and A.~Pakhomov for interesting discussions at the early stage of this
work. A special thank to P.~Strolin for encouragement and advice on the
potentiality of future proton drivers.

\appendix

\section*{Appendix}
\label{sec:appendix}
In the last decade, interest towards the construction of high
intensity proton drivers in the few GeV range has steadily grown. The
main reason is connected with the wide range of applications that can
be simultaneously accessed by these facilities. The most intense
neutron beams (spallation sources) are currently produced by
bombarding mercury targets with energetic protons from a large few-GeV
proton accelerator complex. As it is well known, applications range
from chemistry to crystalline and disordered materials studies,
superconductivity, polymers and structural biology investigations. A
discussion of the phy\-sics case for intense spallation sources can be
found in~\cite{ess_physcase}.  A wide physics program is accessible in
neutrino physics, beyond the oscillation issues discussed
above~\cite{workshopnss}.  The availability of a high power, high duty
factor proton beam could provide opportunities in stopped muon
physics, in search for rare decays as $\mu \rightarrow e\gamma$, $\mu
\rightarrow eee$ or muon conversion $\mu N \rightarrow eN$, and
improvements in muon decay properties. The neutrino energy range is
appropriate for neutrino-nucleus cross section measurements of
relevance to supernova astrophysics: dynamics, nucleosynthesis and
terrestrial supernova $\nu$ detection.  Similarly, measurements of
neutrino-nucleus cross sections open the possibility to study
interesting nuclear structure issues related to the weak interaction
as the ratio of the axial to vector coupling constants and the search
for non-standard contributions.  Finally it's worth noting that
oscillation studies at the $\Delta m^2 \sim 1$~eV$^2$ scale will
become a major priority in $\nu$ physics in case of confirmation of
the LSND $\nu_e$ appearance claim.

%

\end{document}